\documentclass{emulateapj}

\newcommand{\lsun}{\mbox{L$_\odot$}}% Lsun
\newcommand{\msun}{\mbox{M$_\odot$}}% Msun
\newcommand{\rsun}{\mbox{R$_\odot$}}% Rsun

\newcommand{\teff}{\mbox{$T_{eff}$}} % effective temperature
\newcommand{\menv}{\mbox{$M_{env}$}} % 
\newcommand{\lint}{\mbox{$L_{int}$}} % 
\newcommand{\mcore}{\mbox{$M_{core}$}} % 
\newcommand{\rcent}{\mbox{$R_{cent}$}} % 
\newcommand{\rout}{\mbox{$R_{out}$}} % 
\newcommand{\ang}{\mbox{$\alpha$}} %\mbox{$\alpha$}} % 
\newcommand{\incl}{\mbox{$i$}} %\mbox{$i$}} % 

\shorttitle{A Candidate FHSC}
\shortauthors{Enoch et al.}

\begin{document}

\title{A Candidate Detection of the First Hydrostatic Core}

\author{Melissa L. Enoch (1,2), Jeong-Eun Lee (3), Paul Harvey (4), Michael M. Dunham (4), and Scott Schnee (5)}

\affil{
(1) Department of Astronomy, University of California at Berkeley, 601 Campbell Hall, Berkeley, CA, 94720 \\
(2) \textit{Spitzer} Fellow; menoch@berkeley.edu \\
(3) Department of Astronomy and Space Science, Astrophysical Research Center for the Structure and Evolution of the Cosmos, Sejong University, Seoul 143-747 Korea \\
(4) Department of Astronomy, The University of Texas at Austin, 1 University Station, C1400, Austin, TX 78712 \\
(5) National Research Council Canada, Herzberg Institute of Astrophysics, 5071 West Saanich Road, Victoria, BC V9E 2E7,Canada
}

\begin{abstract}
The first hydrostatic core (FHSC) represents a very early phase in the low-mass star formation process, after collapse of the parent core has begun but before a true protostar has formed.  This large (few AU), cool (100 K), pressure supported core of molecular hydrogen is expected from theory, but has yet to be observationally verified.
Here we present observations of an excellent candidate for the FHSC phase: Per-Bolo~58, a dense core in Perseus that was previously believed to be starless.  The $70~\micron$ flux of 65~mJy, from new deep \textit{Spitzer} MIPS observations, is consistent with that expected for the FHSC.  
A low signal-to-noise detection at $24~\micron$ leaves open the possibility that Per-Bolo~58 could be a very low luminosity protostar, however.  We utilize radiative transfer models to determine the best-fitting FHSC and protostar models to the spectral energy distribution and 2.9~mm visibilities of Per-Bolo~58.  The source is consistent with a FHSC with some source of lower opacity through the envelope allowing $24~\micron$ emission to escape; a small outflow cavity or a cavity in the envelope are both possible.  While we are unable to rule out the presence of a protostar, if present it would be one of the lowest luminosity protostellar objects yet observed, with an internal luminosity of $\sim 0.01 \lsun$.
\end{abstract}

\section{Introduction}

One stage of the low mass star formation paradigm has yet to be observed: the first core or first hydrostatic core (FHSC).  
During this brief phase between the initiation of core collapse and the appearance of a ``true'' protostar, the growing central core of molecular hydrogen (H$_2$) is heated to the point at which it obtains hydrostatic balance.  When the central temperature of the first core reaches approximately 2000~K, the H$_2$ is dissociated and a second collapse forms the true protostar.

The FHSC has long been expected from theory \citep{larson69}, 
but has not yet been observationally verified.  \citet{chen10} recently presented a promising candidate, L1448 IRS2E, based on detection of a CO outflow without a corresponding mid-infrared source.  The observed outflow is faster than would be expected for a typical FHSC \citep{mach08}, but might be consistent with the very end of the FHSC phase. 

The expected effective temperature of the FHSC, which has a radius of several AU, is approximately 100~K; as the majority of this cool emission is absorbed and re-radiated by the even cooler envelope, the emergent spectrum of the FHSC is very similar to that of a starless core.  The strongest observational signature is likely to be an increase in luminosity for $\lambda\sim 40-100~ \micron$ as compared to starless cores \citep{by95}.  
Furthermore, given the expected short lifetime of the FHSC ($t\sim10^3 - 3\times10^4$ yr; \citealt{omukai07,by95}), there should only be a few observable even in large samples of cores and protostars.  Based on the relative lifetimes of the FHSC and true protostars ($5.4\times10^5$ yr; \citealt{evans09}), we expect only one FHSC for every 18 to 540 protostars.
Before large \textit{Spitzer} surveys such as the ``From Molecular Cores to Planet-forming Disks'' Legacy Program (``Cores to Disks'' or c2d; \citealt{evans03}), the number of known protostars in any given region was typically too low to expect even one FHSC.  

We have recently completed a census of starless cores and embedded protostars in the Perseus, Serpens, and Ophiuchus molecular clouds \citep{enoch08,enoch09a}, based on large-scale Bolocam 1.1~mm continuum surveys \citep{enoch06,young06,enoch07} and IRAC and MIPS maps from c2d \citep{jorg06,harv06,reb07,harv07a,padgett08}.
Perseus, Serpens, and Ophiuchus each harbor between 30 and 70 embedded protostars -- objects after the second collapse, and thus more evolved than FHSCs, but still embedded in and accreting from their natal cores -- and should contain at least a few FHSCs (if the lifetime is longer than $10^4$ yr). 
\textit{Spitzer} was sufficiently sensitive to detect $70~\micron$ emission from a 0.015~\msun\ FHSC at a distance of 250~pc \citep{omukai07}, but large surveys such as c2d were not deep enough at $70~\micron$ to detect the FHSC.

In this Letter, we present new deep \textit{Spitzer} MIPS observations that reveal a candidate FHSC in the Perseus molecular cloud.  
Utilizing literature data, we compare the spectral energy distribution (SED) and millimeter visibilities of this source to radiative transfer models of a FHSC and very low luminosity protostar.

\section{Target Selection}

\begin{figure}
\plotone{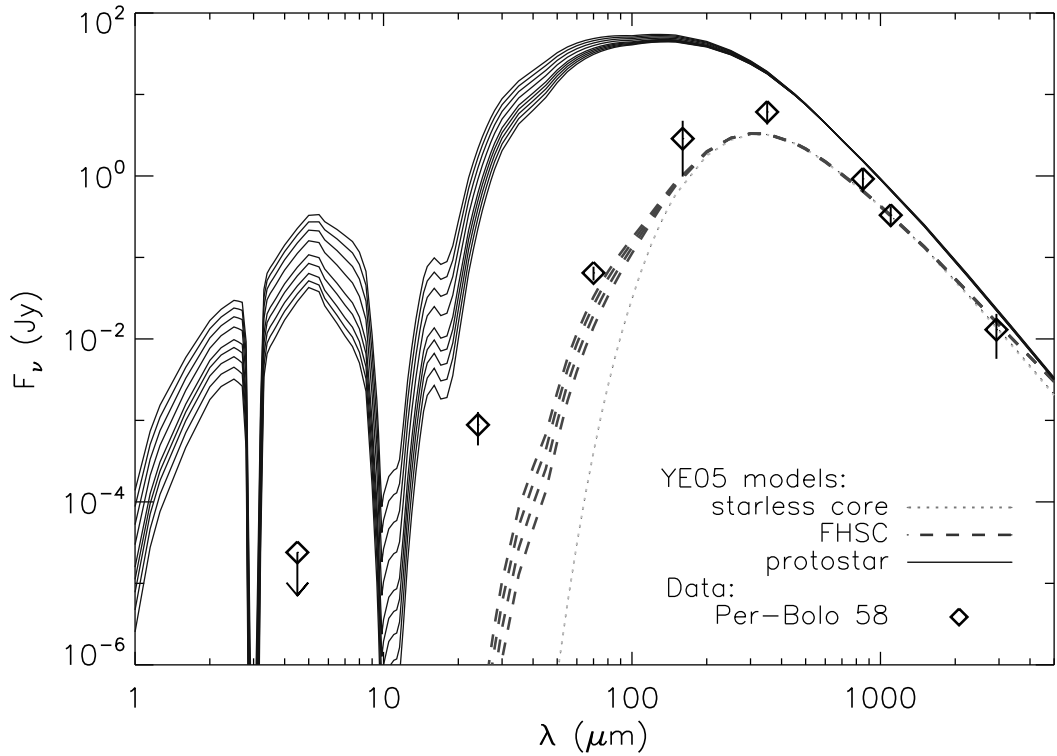}
\caption
{
Radiative transfer models from \citet{ye05}, for a $1~\msun$ core at the distance of Perseus ($d=250$~pc).  SEDs represent a time sequence, from a starless core (dotted line) to FHSC (dashed lines) to true protostar (solid lines), with short wavelength flux increasing monotonically with time and time steps of 2000~yr.  The observed SED of Per-Bolo 58 (diamonds; see Section~\ref{ressec}) includes fluxes from Table~\ref{fluxtab}.
The $70~\micron$ flux of Per-Bolo~58 is roughly consistent with that expected for the FHSC.  The $24~\micron$ detection is inconsistent with these simple spherical models of the FHSC, however. 
\label{predictfig}}
\end{figure}

Figure~\ref{predictfig} shows the expected observed SED of a FHSC, from the radiative transfer models of \citet{ye05}.
Based on these models, an unambiguous detection of the FHSC requires detection of compact $70~\micron$ emission in a dense core with no emission at $\lambda \lesssim 24~\micron$.  
The $70 ~\micron$ sensitivity limit of the c2d survey was approximately 75 mJy ($5\sigma$), 
somewhat higher than the flux expected for a typical FHSC \citep{by95,ye05}.
Thus any FHSCs appeared as starless cores in our census, and deeper $70~\micron$ follow-up observations were necessary to reveal them.\footnote{We did an initial check of the c2d catalogs for starless cores detected only at $70~\micron$ but found no such objects.}

We selected targets from our census of starless cores \citep{enoch08}, which includes 108 cores and is complete to $\mcore \gtrsim 0.2 \msun$, based on three criteria. (1) Visual inspection of the c2d $70~\micron$ maps at starless core positions, looking for faint emission below the formal $5\sigma$ detection limit.  (2) Starless cores with mass-to-size ratios above the critical value for stable Bonnor-Ebert spheres 
(central to surface density ratio $\rho_c / \rho_0 > 14$, or M [$\msun$]/FWHM [pc] $> 31.9$ for $T=10$~K).  
These cores are likely to be collapsing, but the lack of near- or mid-infrared emission indicates that they have not yet formed a true protostar.
(3) Spatially compact starless cores, either unresolved by the $31\arcsec$ Bolocam beam or compact in follow-up CARMA 2.9~mm interferometric maps (Enoch et al., in prep).  Again, these cores have high central density and may be collapsing. 

In the three clouds, 7 starless cores met one or more of the above criteria.
This is by no means a complete or unbiased sample; rather, we used these criteria to identify interesting objects for deeper follow-up observations.

\section{Observations and Results}\label{ressec}

Deep $70~\micron$ maps of our 7 targets were obtained with the Multi-Band Imaging Photometer for Spitzer (MIPS) on the \textit{Spitzer Space Telescope} \citep{werner04} during 2008 October 20 - October 27.  
Integration times ranged between $1100-2000$ s pixel$^{-1}$, depending on the brightness of the background estimated from c2d maps.  The resulting $5\sigma$ detection limit of $\sim 7.5$~mJy was chosen based on the models of \citet{omukai07}, to detect FHSCs with mass $\gtrsim 0.015~\msun$ at the distance of Perseus and Serpens ($250-260$~pc).  

\begin{figure}
\plotone{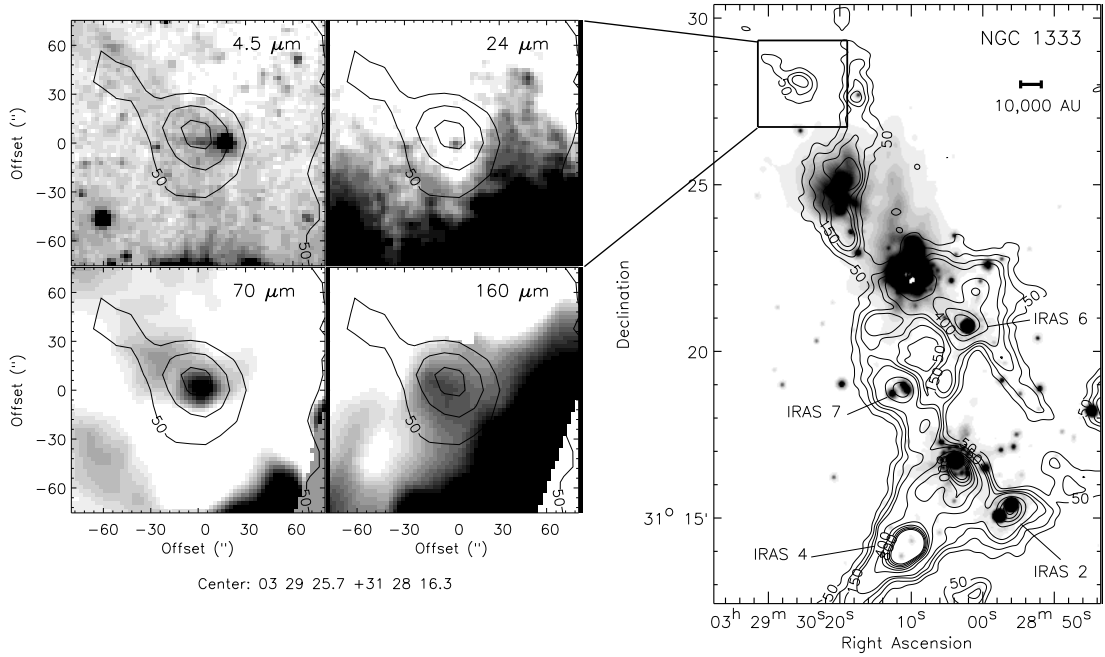}
\caption
{\textit{Spitzer} images of Per-Bolo~58 at $4.5, 24, 70$ and $160~\micron$ (left).  Bolocam 1.1~mm continuum contours indicate the dense core, previously believed to be starless.  The $70~\micron$ map is from our new deep observations; other images are from c2d.  
At right is a $24~\micron$ (grayscale) and 1.1~mm (contours) map of the surrounding region in NGC~1333.  
1.1~mm contours are $50,100...200,400...1400$~mJy beam$^{-1}$ in both panels, and IRAS sources indicated are from \citet{jenn87}.
\label{mapfig}}
\end{figure}

Data were reduced using the c2d pipeline \citep{evans07}.
One source was clearly detected at $70~\micron$: [EYG2006] Bolo~58 (RA=03 29 25.7, Dec=+31 28 16.3), hereafter Per-Bolo~58, a dense core identified as starless by both \citet{enoch06} and \citet{hatch07}. 
Per-Bolo~58 was included in our target list based on faint emission in the c2d $70~\micron$ map and a spatially compact 1.1~mm core.  It is located in the outskirts of the NGC~1333 young cluster (see Figure~\ref{mapfig}, right).

The $70~\micron$ flux of Per-Bolo~58 is $65\pm6$~mJy.
A subsequent search of the c2d deep catalogs \citep{evans07} revealed a low signal-to-noise $24~\micron$ detection at the position of Per-Bolo~58, with a flux of $0.88\pm0.24$~mJy.
This source was not included in the ``high reliability'' or ``YSO candidate'' c2d catalogs, as it did not meet the $7\sigma$ signal-to-noise requirement. 
Given the good positional correspondence and point-like emission, we believe the $24~\micron$ source is associated with Per-Bolo~58.  There is also a nearby low signal-to-noise $4.5~\micron$ detection ($0.024\pm0.007$ mJy) that we do not believe is associated, as the only point-like emission is offset by $10\arcsec$ from the $24~\micron$ and $70~\micron$ sources.  We treat the $4.5~\micron$ point as an upper limit.
We estimate a $160~\micron$ flux of 2.8~Jy from the c2d MIPS $160~\micron$ map; this value has a large uncertainty ($\sim50$\%) due to non-uniform extended emission in NGC~1333.  
\textit{Spitzer} images at $4.5, 24, 70$, and $160~\micron$ are shown in Figure~\ref{mapfig}. 

All known continuum fluxes for Per-Bolo~58 are given in Table~\ref{fluxtab}, and the observed SED is shown in Figure~\ref{predictfig}.
In addition to the \textit{Spitzer} data described above, we include SHARC II $350~\micron$ (Dunham et al., in prep), SCUBA $850~\micron$ \citep{hatch05}, and Bolocam 1.1~mm \citep{enoch06} fluxes, as well as the CARMA 2.9~mm map from \citet{schnee10}.  
The interferometric CARMA 2.9~mm map provides both a long wavelength flux measurement and a measure of the resolved radial intensity profile.
The 2.9~mm visibility amplitudes versus $uv$-distance are shown in Figures~\ref{fit1fig} and \ref{fit2fig}.

\section{Discussion}\label{discsec}

Our $70~\micron$ detection of Per-Bolo~58 indicates the presence of an internal luminosity source in this dense core, with a flux consistent with that expected for a FHSC.
The detection at $24~\micron$ is inconsistent with a FHSC surrounded by a spherically symmetric envelope, however (Figure~\ref{predictfig}), leaving open the possibility that Per-Bolo~58 may be a very low luminosity protostar.  
We discuss these possibilities below. 

\subsection{Radiative Transfer Models}\label{modsec}

We use RADMC, a two-dimensional Monte Carlo radiative transfer code \citep{dd04} to model a FHSC and low luminosity protostar.
We adopt a density profile that includes a rotating infalling envelope characterized by total mass \menv, centrifugal radius \rcent, and outer radius \rout\ \citep[e.g.][]{ulrich67}, and outflow cavity characterized by full opening angle $\ang$.  
The envelope mass is held fixed at $\menv=1.2~ \msun$, as calculated from the 1.1~mm flux (see \citealt{enoch06}) assuming a dust temperature of $T=8$~K, consistent with radiative transfer models of such a low luminosity source. 
Details of the density profile and modeling are as in \citet{enoch09b}.

Dust opacities are from Table 1, column 5 of
\citet{oh94} for dust grains with thin ice mantles, including scattering.
Following \citet{dunham10}, we remove the effects of excess back-scattering of the interstellar radiation field (ISRF) by subtracting the flux for a starless core model at short wavelengths.  \citet{dunham10} make this correction for $\lambda\le10~\micron$; here we additionally include the small plateau of ISRF emission at $\lambda=10-40 ~\micron$.  While this correction affects the shape of our model SEDs at short wavelengths, the correction at $24~\micron$ is small, so it does not significantly affect the model fits. 

To match the peak of the SED at $80~\micron < \lambda < 1000~\micron$ we include an ISRF that is a factor of 3 stronger than the ``Black-Draine'' ISRF \citep{evans01,black94,draine78}.  
A more intense radiation field may be expected in NGC 1333, where there is significant extended emission at $\lambda \le 24 ~\micron$ (e.g. Figure~\ref{mapfig}).  
Increasing the ISRF improves the overall fit of all models, but does not change the best-fitting envelope parameters, because the SED peak depends primarily on fixed parameters (\menv, ISRF).

\clearpage
\subsection{Per-Bolo~58 as a First Hydrostatic Core}\label{fhscsec}

Our FHSC models include a cool ($\teff=100$~K), large ($R\sim 2$~AU, where the actual radius is determined from $\lint=4\pi R^2 \sigma \teff^4$) internal luminosity source.
We run a grid of models varying \lint\ (0.006, 0.012, 0.025 \lsun), \rout\ (4000, 6000, 8000, 10000~AU), \rcent\ (10, 50, 100, 200, 300~AU), \ang\ (0, 5, 10, 20, 40, 60~$\deg$), and inclination (\incl; 5, 10, 15...90~$\deg$).  

Models are compared to the observed SED of Per-Bolo~58 using a $\chi^2$ analysis to constrain \lint, \rcent, \ang, and \incl.  
An internal luminosity of $0.012~\lsun$ is clearly preferred by the SED, and we derive other parameters with \lint\ fixed.  The SED is rather insensitive to \rout, so we also compute the visibility amplitudes of each model as a function of $uv$-distance, and compare to the 2.9~mm CARMA observations from \citet{schnee10}.
At small $uv$-distances ($\le 10 k\lambda$), the visibilities are sensitive to the spatial extent of the envelope (see \citet{enoch09b} for more details), and we find a best fit for $\rout=8000$~AU.

\begin{figure}
\plotone{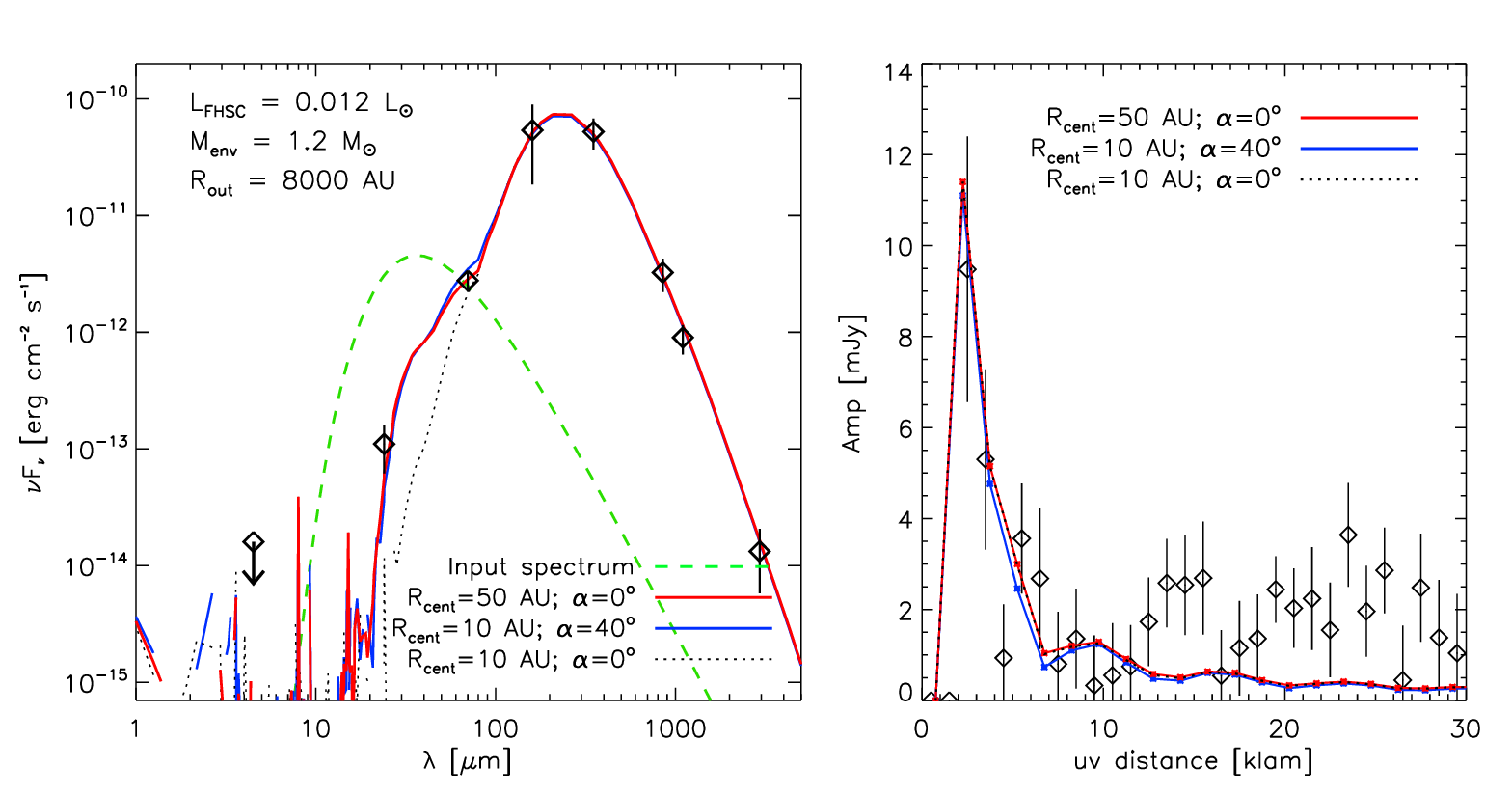}
\caption
{Best-fitting FHSC model compared to the observed SED (left) and CARMA 2.9~mm visibilities (right; \citealt{schnee10}) of Per-Bolo~58. While a FHSC  provides a good fit to the SED, either a cavity in the inner envelope ($\rcent \sim 50$~AU), or outflow ($\ang \gtrsim 20 \deg$) is required to match the $24~\micron$ flux.
The two best models, with $\rcent=50$~AU, $\ang=0 \deg$ (red) and $\rcent=10$~AU, $\ang=40 \deg$ (blue) are nearly indistinguishable.  
The input FHSC spectrum ($L=0.012~\lsun$, $R=1.7$~AU, $\teff=100$~K; dashed line), and a model with no cavity (dotted line) are also shown for comparison.  The millimeter visibilities help to constrain the envelope outer radius, with a best fit for $\rout=8000$~AU. 
\label{fit1fig}}
\end{figure}

\begin{figure}
\plotone{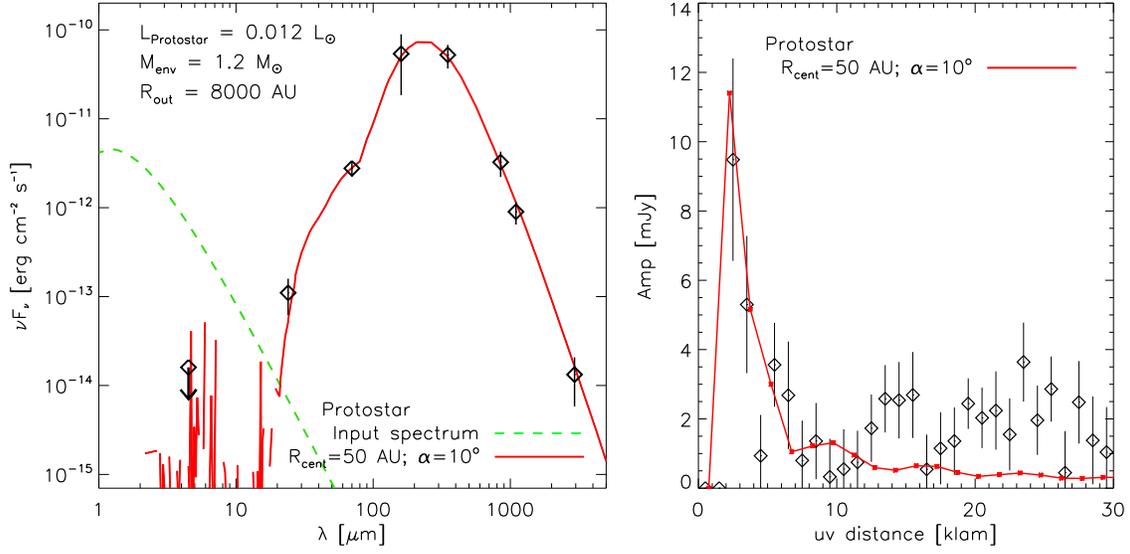}
\caption
{Same as Figure~\ref{fit1fig}, but for the best-fitting protostar model, with $\lint=0.012~\lsun$ and a narrow outflow cavity ($\ang\sim10 \deg$).  
Despite the much hotter input spectrum (3000~K), the best-fitting protostar SEDs is nearly indistinguishable from the best-fitting FHSC, because most of the internal luminosity is re-processed by the dense envelope.  
Likewise, the 2.9~mm visibilities depend only on the envelope density profile, which is similar in both sets of models. 
\label{fit2fig}}
\end{figure}

The best fitting FHSC model ($\lint=0.012~\lsun$, $\teff=100$~K, $R=1.7$~AU, $\rcent=50$~AU, $\rout=8000$~AU, $\ang=0~\deg$, $\incl=25~\deg$) is compared to the observed SED and 2.9~mm visibilities in Figure~\ref{fit1fig}. Another well-fitting model with $\rcent=10$~AU and $\ang=40\deg$ is also shown.  A number of models have similar $\chi^2$ values, but all have either an envelope cavity ($\rcent \ge 50$~AU) or an outflow ($\ang \ge 20~\deg$). 
Models without an envelope cavity are clearly unable to match the observed $24~\micron$ point, due to the high opacity through the envelope; such a model with $\rcent=10$~AU and $\ang=0~\deg$ is shown in Figure~\ref{fit1fig} for comparison.

Although an envelope cavity is not expected from simple physical models of the FHSC, in which the envelope extends down to the FHSC and no outflow is present, there are several cases in which we might expect a cavity.  
MHD calculations have shown that the FHSC can drive a molecular outflow \citep[e.g.][]{mach08}.  Such early outflows should be weak, but could evacuate enough envelope material to allow a small amount of $24~\micron$ flux to escape.  Similarly, a larger centrifugal radius might be expected if the inner envelope were cleared by a binary source or weak outflow.  We note that $4.5~\micron$ is the \textit{Spitzer} band associated with shock emission in outflows, and the nearby weak $4.5~\micron$ emission could be tracing a weak outflow.

\subsection{Per-Bolo~58 as a Very Low Luminosity Protostar}\label{vellosec}

A number of dense cores previously believed to be starless were found by \textit{Spitzer} to harbor very low luminosity objects (VeLLOs; e.g. \citealt{young04}).
The best current explanation of VeLLOs is that they are true protostars, but with low internal luminosities due either to very low masses or very low accretion rates.
Most VeLLOs have internal luminosities between $0.02$ and $0.1~\lsun$ \citep{dun08}; although the $24$ and $70~\micron$ fluxes of Per-Bolo~58 are lower than typical VeLLOs, it may be an extreme example of a low luminosity protostellar population.  At least two other protostellar sources with similar luminosities are known: a candidate embedded proto-brown dwarf in Taurus with $\lint\sim0.003~ \lsun$ \citep[SSTB213 J041757;][]{barrado09}, and Cha-MMS~1, which has an estimated internal luminosity of $0.01-0.02~\lsun$ \citep{belloche06}\footnote{\citet{belloche06} note that Cha-MMS~1 may also be a FHSC candidate, however.}.

To model Per-Bolo~58 as a protostar, we use a similar grid as above, but include a hot internal luminosity source typical of a true protostar ($\teff=3000$~K, $R \sim 1~\rsun$).  The best fitting protostar model ($\lint=0.012~\lsun$, $\rcent=50$~AU, $\rout=8000$~AU, $\ang=10~\deg$, $\incl=30~\deg$) is shown in Figure~\ref{fit2fig}.  Protostar models with relatively high inclinations ($\incl > \ang$) are nearly indistinguishable from the corresponding FHSC models; in both cases essentially all of the emission from the central source is reprocessed by the envelope, masking the details of the internal luminosity source.  We can rule out protostar models with low inclinations, however, which produce far too much emission in the near-infrared.

Again, an internal luminosity of 0.012~\lsun is strongly favored.  This value is corroborated by the relationship between internal luminosity and observed $70~\micron$ flux determined by \citet{dun08}: $L_{\mathrm{int}} = 3.3 \times10^8 F_{70}^{0.94}~ \lsun$, where $F_{70}$ is the flux (in erg~cm$^{-2}$~s$^{-1}$) at 140~pc.  For Per-Bolo~58 this relationship yields $L_{\mathrm{int}} = 0.014~ \lsun$.
If Per-Bolo~58 is a true protostar then it is one of the lowest luminosity embedded protostars known, with $\lint\sim0.01~\lsun$.

\subsection{Possible Observational Tests}\label{futuresec}

We are currently unable to conclusively determine the evolutionary state of Per-Bolo~58.  
While the Herschel Gould Belt survey \citep{andre10} will further refine the SED, additional observations are needed to  distinguish  a FHSC with a small envelope cavity from an extremely low luminosity protostar.
Possible observational tests include measuring the outflow velocity and finding evidence of water ice evaporation.

Numerical MHD models suggest that the FHSC can drive a molecular outflow, but with quite low velocity ($v\sim3$~km~s$^{-1}$; \citealt{mach08,tomida10}).  As protostellar outflows typically have speeds of $10-20$~km~s$^{-1}$ \citep[e.g.][]{as06}, detecting an outflow and measuring its velocity, either in the millimeter or with Herschel spectroscopy, could test our hypothesis that Per-Bolo~58 is a FHSC.
No outflow was detected in the HARP $^{12}$CO (3-2) study of \citet{hd09}, but a FHSC outflow should be below their detection criteria (linewing strength above 1.5~K at 3~km~s$^{-1}$ from the core velocity).  

Unlike the FHSC, for which the maximum dust temperature in the envelope is only $\sim80$~K, temperatures in our protostar models are high enough in the inner few AU (up to 300~K) to evaporate water ice \citep[e.g.][]{fraser01}.  ALMA is capable of detecting the resulting H$_2$O or HDO lines, around 300~GHz and 240~GHz, even with the beam dilution of an $0.08\arcsec$ (20~AU at 250~pc) beam.  
Given the outflow non-detection at the level expected for a typical protostar, finding chemical signatures indicative of high gas temperatures may be the best way to rule out Per-Bolo~58 as a FHSC.

\section{Conclusions}\label{sumsec}
In an effort to observationally verify the theoretically predicted FHSC phase, we have obtained deep \textit{Spitzer} $70~\micron$ images of a small sample of dense starless cores.
Per-Bolo~58 is detected at the level expected for the FHSC, making it a very promising FHSC candidate. A weak detection at $24~\micron$ leaves open the possibility that it might be an extremely low luminosity protostar, however. 

We are able to reproduce the observed SED and 2.9 mm visibilities of Per-Bolo~58 with radiative transfer models of a FHSC, although some source of lower envelope opacity allowing $24~\micron$ emission to escape is required: either a small outflow cavity or a spherical cavity in the envelope of $\sim50$~AU.
While we cannot rule out the possibility that Per-Bolo~58 has already formed a true protostar, with an internal luminosity $\lint\sim0.01~ \lsun$ it would be one of the lowest luminosity protostars known.
Additional observational tests to clarify the evolutionary state of Per-Bolo~58 include measuring the velocity of any outflow present, and looking for evidence of water ice evaporation in the inner few AU. 
Finally, we note that if Per-Bolo~58 is a true FHSC, then its observational similarity to VeLLOs (which are defined by luminosity alone) suggests that some VeLLOs currently thought to be protostellar might in fact be members of the FHSC phase.

\acknowledgments
We thank the anonymous referee for helpful comments, C. Dullemond for the use of RADMC, and N. Evans for many useful discussions.
This work is based on observations made with the Spitzer Space Telescope, operated by the Jet Propulsion Laboratory (JPL), California Institute of Technology (Caltech) under a contract with NASA. Support was provided by NASA through an award issued by JPL/Caltech, by the Spitzer Fellowship Program, and by the National Research Foundation of Korea (NRF) grant funded by the Korea government (No. 2009-0062866) and Basic Science Research Program through the NRF funded by the Ministry of Education, Science and Technology (No. 2010-0008704).

\clearpage

\begin{deluxetable}{ccccl}
\tablecolumns{5}
\tablewidth{0pc}
\tablecaption{\label{fluxtab} Observed SED of Per-Bolo 58}
\tablehead{
\colhead{Wavelength}  & \colhead{Flux} & \colhead{$\sigma$(Flux)} & \colhead{Aperture} & \colhead{Notes} \\
\colhead{($\micron$)}  & \colhead{(mJy)} & \colhead{(mJy)} & \colhead{diameter ($\arcsec$)} & \colhead{}
}
\startdata
       4.5  &  0.024  & 0.07  & $2.2$ &   upper limit; Spitzer (c2d)    \\
       24   &  0.88  &  0.24  & $7$ &   Spitzer (c2d)    \\
       70   &  65  &   6      & $17$ &    Spitzer; this work   \\
       160  &  2870  &  1600  & $40$ &    Spitzer (c2d)   \\
       350  &  6100  &  1200  & $40$ &   SHARC II; Dunham et al., in prep    \\
       850  &  920  &  200    & $18$ &   SCUBA; \citet{hatch05}   \\
       1100 &  330  &  30     & $40$ &  Bolocam; \citet{enoch06}      \\
       2930 &  13    &  6     & $15$ &  CARMA; \citet{schnee10}      
\enddata
\tablecomments{Continuum fluxes are calculated by aperture photometry within the aperture listed, with the exception of the 2.9~mm flux, which is determined by a Gaussian fit (FWHM$=15\arcsec$).  Our 2.9~mm flux does not include the SZA data from \citet{schnee10}, which probes larger spatial scales than we include in our models.} 
\end{deluxetable}

\begin{thebibliography}{}
\bibitem[Andr\'{e} et al.(2010)]{andre10} Andr\'{e}, P., et al. 2010, \aap, 518, 102
\bibitem[Arce \& Sargent(2006)]{as06} Arce, H. G. \& Sargent, A. I. 2006, \apj, 646, 1070
\bibitem[Barrado et al.(2009)]{barrado09} Barrado, D., Morales-Calder\'{o}n, M., Palau, A., Bayo, A., de Gregorio-Monsalvo, I., Eiroa, C.; Hu\'{e}lamo, N., Bouy, H., Morata, O., \& Schmidtobreick, L.  2009, \aap, 508, 859
\bibitem[Belloche et al.(2006)]{belloche06} Belloche, A., Parise, B., van der Tak, F. F. S., Schilke, P., Leurini, S., G\"{u}sten, R., \& Nyman, L.-\AA. 2006, \aap, 454, L51
\bibitem[Black(1994)]{black94} Black, J. H. 1994, ASP Conf. Ser. 58, The First Symposium on the Infrared Cirrus and Diffuse Interstellar Clouds, ed. R. M. Cutri \& W. B. Latter (San Francisco: ASP), 355
\bibitem[Boss \& Yorke(1995)]{by95} Boss, A. P \& Yorke, H. W.  1995, \apj, 439L, 55
\bibitem[Chen et al.(2010)]{chen10} Chen, X., Arce, H. G., Zhang, Q., Bourke, T. L., Launhardt, R., Schnalzl, M., \& Henning, T. 2010 \apj, 715, 1344
\bibitem[Draine(1978)]{draine78} Draine, B. T. 1978, \apjs, 36, 595 
\bibitem[Dunham et al.(2008)]{dun08} Dunham, M. M., Crapsi, A., Evans, N. J., II, Bourke, T. L., Huard, T. L., Myers, P. C., \& Kauffmann, J.  2008, \apjs, 179, 249
\bibitem[Dunham et al.(2010)]{dunham10} Dunham, M., Evans, N. J., Terebey, S., Dullemond, C., P., \& Young, C. H. 2010, \apj, 710, 470
\bibitem[Dullemond \& Dominik(2004)]{dd04} Dullemond, C. P. \& Dominik, C.  2004, \aap, 417, 159
\bibitem[Enoch et al.(2006)]{enoch06} Enoch, M. L., Young, K. E., Glenn, J., Evans, N. J., II, Golwala, S., Sargent, A. I., Harvey, P., et al. 2006, \apj, 638, 293
\bibitem[Enoch et al.(2007)]{enoch07} Enoch, M. L., Glenn, J., Evans, N. J., II, Sargent, A. I., Young, K. E., \& Huard, T. L.  2007, \apj, 666, 982
\bibitem[Enoch et al.(2008)]{enoch08} Enoch, M. L., Evans, N. J., II, Sargent, A. I., Glenn, J., Rosolowsky, E., \& Myers, P. C.  2008, \apj, 684, 1240
\bibitem[Enoch et al.(2009a)]{enoch09a} Enoch, M. L., Evans, N. J., II, Sargent, A. I., Glenn, J. 2009, \apj, 692, 973
\bibitem[Enoch et al.(2009b)]{enoch09b} Enoch, M. L., Corder, S., Dunham, M. M, Duch\^{e}ne, G. 2009, \apj, 707, 103
\bibitem[Evans et al.(2001)]{evans01} Evans, N. J., II, Rawlinge, J. M. C., Shirley, Y. L., \& Mundy, L. G.  2001, \apj, 557, 193
\bibitem[Evans et al.(2003)]{evans03} Evans, N. J., II, et al. 2003, \pasp, 115, 965
\bibitem[Evans et al.(2007)]{evans07} Evans, N. J., II, et al. 2007, Final Delivery of Data from the c2d Legacy Project: IRAC and MIPS (Pasadena: SSC)
\bibitem[Evans et al.(2009)]{evans09} Evans, N. J., II, et al. 2009, \apjs, 181, 321
\bibitem[Fraser et al.(2001)]{fraser01} Fraser, H. J., Collings, M. P., McCoustra, M. R. S., \& Williams, D. A.  2001, MNRAS, 327, 1165
\bibitem[Harvey et al.(2006)]{harv06} Harvey, P. M., Chapman, N., Lai, S.-P.. Evans, N. J., II, Allen, L. E., J{\o}rgensen, J. K., Mundy, L. G., et al. 2006, \apj, 644, 307
\bibitem[Harvey et al.(2007a)]{harv07b} Harvey, P. M., Merin, B., Huard, T. L., Rebull, L. M., Chapman, N. Evans, N. J., II, \& Myers, P. C. 2007a, \apj, 663, 1149 
\bibitem[Harvey et al.(2007b)]{harv07a} Harvey, P. M., Rebull, L. M., Brooke, T., Spiesman, W. J., Chapman, N., Huard, T. L., Evans, N. J., II, et al. 2007b, \apj, 663, 1139 
\bibitem[Hatchell et al.(2005)]{hatch05} Hatchell, J., Richer, J. S., Fuller, G. A., Qualtrough, C. J., Ladd, E. F., \& Chandler, C. J.  2005, \aap, 440, 151
\bibitem[Hatchell et al.(2007)]{hatch07} Hatchell, J., Fuller, G. A., Richer, J. S., Harries, T. J., \& Ladd, E. F.  2007, \aap, 468, 1009
\bibitem[Hatchell \& Dunham(2009)]{hd09} Hatchell, J. \& Dunham, M. M.  2009, \aap, 502, 139
\bibitem[Jennings et al.(1987)]{jenn87} Jennings, R. E., Cameron, D. H. M., Cudlip, W., \& Hirst, C. J.  1987, \mnras, 226, 461
\bibitem[J{\o}rgensen et al.(2006)]{jorg06} J{\o}rgensen, J. K., Harvey, P. M., Evans, N. J., II, Huard, T. L., Allen, L. E., Porras, A., Blake, G. A., et al. 2006, \apj, 645, 1246
\bibitem[Larson(1969)]{larson69} Larson, R. B.  1969, \mnras, 145, 271
\bibitem[Machida et al.(2008)]{mach08} Machida, M., Inutsuka, S.-I., \& Matsumoto, T. 2008, \apj, 676, 1088 
\bibitem[Omukai(2007)]{omukai07} Omukai, K. 2007, PASJ, 59, 589 
\bibitem[Ossenkopf \& Henning(1994)]{oh94} Ossenkopf, V., \& Henning, Th.  1994, \aap, 291, 943 
\bibitem[Padgett et al.(2008)]{padgett08} Padgett, D. L., et al. 2008, \apj, 672, 1013
\bibitem[Rebull et al.(2007)]{reb07} Rebull, L. M., et al.  2007, \apjs, 171 447
\bibitem[Schnee et al.(2010)]{schnee10} Schnee, S., Enoch, M., Johnstone, D., Culverhouse, T., Leitch, E., Marrone, D. P., \& Sargent, A. 2010, \apj, 718, 306
\bibitem[Tomida et al.(2010)]{tomida10} Tomida, K., Tomisaka, K., Matsumoto, T., Ohsuga, K., Machida, M. N., \& Saigo, K.  2010 \apj, 714, L58
\bibitem[Ulrich et al.(1967)]{ulrich67} Ulrich, B. T., Neugebauer, G., McCammon, D., Leighton, R. B., Hughes, E. E., \& Becklin, E.  1967, \apj, 147, 858
\bibitem[Werner et al.(2004)]{werner04} Werner, M., et al. 2004, \apjs, 154, 1 
\bibitem[Young \& Evans(2005)]{ye05} Young, C. H. \& Evans, N. J., II  2005, \apj, 627, 293
\bibitem[Young et al.(2004)]{young04} Young, C. H., et al. 2004, ApJS, 154, 396
\bibitem[Young et al.(2006)]{young06} Young, K. E., Enoch, M. L., Evans, N. J., II, Glenn, J., Sargent, A., Huard, T. L., Aguirre, J., et al. 2006, \apj, 644, 326

\end{thebibliography}
\end{document}